\documentclass[reprint,aps,prl,superscriptaddress]{revtex4-1}
\usepackage{amsmath}
\usepackage[pdftex]{graphicx}
\usepackage{ulem}

\newcommand{\Pone}{P$\mathrm{_1}$}
\newcommand{\Ptwo}{P$\mathrm{_2}$}
\newcommand{\VGone}{$V_{g1}$}
\newcommand{\VGtwo}{$V_{g2}$}
\newcommand{\VD}{$V_d$}
\newcommand{\VBG}{$V\mathrm{_{bg}}$}
\newcommand{\IDS}{$I_{ds}$}
\newcommand{\GS}{$\Gamma_{s}$}
\newcommand{\GD}{$\Gamma_{d}$}
\newcommand{\GIN}{$\mathrm{\Gamma_{in}}$}
\newcommand{\EConetwo}{$E_{C12}$}
\newcommand{\ZIMP}{z$_{\mathrm{imp}}$}

\begin{document}
\title{Detection of a large valley-orbit splitting in silicon with two-donor spectroscopy}

\author{B. Roche}
\author{E. Dupont-Ferrier}
\author{B. Voisin}
\affiliation{SPSMS, UMR-E CEA / UJF-Grenoble 1, INAC, 17 rue des Martyrs, 38054 Grenoble, France}
\author{M. Cobian}
\affiliation{SPMM, UMR-E CEA / UJF-Grenoble 1, INAC, 17 rue des Martyrs, 38054 Grenoble, France}
\author{X. Jehl}
\affiliation{SPSMS, UMR-E CEA / UJF-Grenoble 1, INAC, 17 rue des Martyrs, 38054 Grenoble, France}
\author{R. Wacquez}
\affiliation{SPSMS, UMR-E CEA / UJF-Grenoble 1, INAC, 17 rue des Martyrs, 38054 Grenoble, France}
\affiliation{CEA, LETI, MINATEC Campus, 17 rue des Martyrs, 38054 Grenoble, France}
\author{M. Vinet}
\affiliation{CEA, LETI, MINATEC Campus, 17 rue des Martyrs, 38054 Grenoble, France}
\author{Y.-M. Niquet}
\affiliation{SPMM, UMR-E CEA / UJF-Grenoble 1, INAC, 17 rue des Martyrs, 38054 Grenoble, France}
\author{M. Sanquer}
\email[]{marc.sanquer@cea.fr}
\affiliation{SPSMS, UMR-E CEA / UJF-Grenoble 1, INAC, 17 rue des Martyrs, 38054 Grenoble, France}

\date{\today}

\begin{abstract}
We measure a large valley-orbit splitting for shallow isolated phosphorus donors in a silicon gated nanowire. This splitting is close to the bulk value and well above previous reports in silicon nanostructures. It was determined using a double dopant transport spectroscopy which eliminates artifacts induced by the environment. Quantitative simulations taking into account the position of the donors with respect to the Si/SiO$_2$ interface and electric field in the wire show that the values found are consistent with the device geometry.
\end{abstract}

\maketitle

Nanofabrication technologies now offer the exciting opportunity to access single dopant features in nanoscale transistors~\cite{Koenraad2011}. Yet the control over two dopants remains challenging. In this letter, we demonstrate electrical transport through two donors in series in a semiconducting nanostructure. By using the electronic ground level of one donor as an energy filter for electrons emitted from the source we measure the valley-orbit splitting (VOS)---the energy separation between the singlet ground and first excited states---of the second donor. Unlike the usual single level spectroscopy, this two-dopant spectroscopy eliminates local density-of-states fluctuations and finite temperature effects in the contacts~\cite{Vaart1995,Pierre2009}. It also helps ruling out the effects of environmental offset charges~\cite{Pierre2009b}.

Using single dopants is a simple way to achieve large single level spacing in silicon. This results from the sharp potential of the impurity, much steeper than in gate-defined quantum dots, which lifts the valley degeneracy of the conduction band. As an example, VOS up to 5\,meV have been reported for an arsenic donor in a silicon device~\cite{Rahman2011}, which is however much lower than the bulk value of 20\,meV. The difference was explained by the presence of a strong electric field which hybridizes the donor and interface orbitals. The VOS of shallow donors can indeed be very dependent on their position in the device~\cite{Rahman2009}. Nearby Si/SiO$_2$ interfaces and electric fields break the symmetry around the impurity and can shift the wave functions away from the donor nucleus, which decreases the VOS.

Here we report a VOS as large as 10\,meV---close to the bulk value of 11.67\,meV~\cite{Ramdas1981}---for a phosphorus atom in a nanometer size silicon device despite the close proximity of gate, source and drain electrodes. This shows unambiguously that large VOS can be reached on dopants in silicon devices without significant perturbation by the environment. To perform this two-level spectroscopy, an independent control over the energy levels of two donors in series is necessary. This is achieved here by the use of two independent front gates. This two-dopant transistor offers multiple possibilities beyond spectroscopy, just like coupled quantum dots compared to single dots.

\begin{figure}[t]
\begin{center}
\includegraphics[width=\columnwidth]{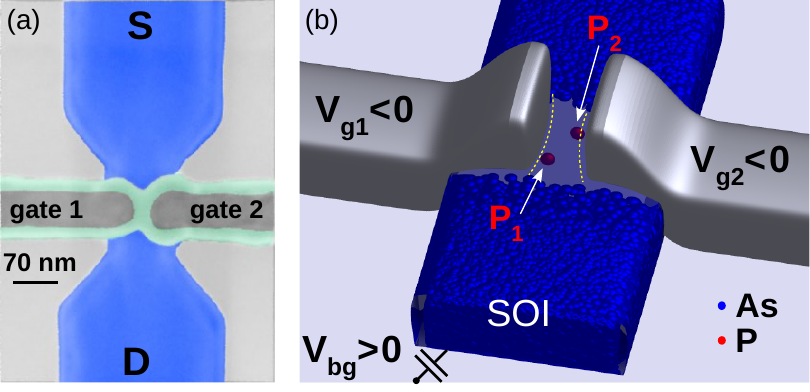}
\caption{(color online). (a) False color top view of the sample (scanning electron microscope). Two gates (gray) surrounded by nitride spacers (green) partially cover the SOI channel between the source (S) and drain (D).
(b) Schematic of the sample with As dopants in the source and drain (spacers not shown for clarity). For positive substrate bias and negative front gate voltages the electrons flow near the buried Si/SiO$_2$ interface in the constriction between the front gates. Two P donors (\Pone{}, \Ptwo{}) are drawn in red in this region.}
\label{fig1}
\end{center}
\end{figure}

The general fabrication technique of our devices can be found in Ref.~\cite{Hofheinz2006}. The 60\,nm wide and 20\,nm thick silicon-on-insulator (SOI) channel is P doped at a concentration of  $10^{18}$ cm$^{-3}$. The nanowire is partially covered by two 40\,nm long polycrystalline silicon front gates on a 5\,nm thick front gate oxide (SiO$_2$), facing each other at a distance of 30\,nm (Fig.~\ref{fig1}).
15\,nm thick Si$_\mathrm 3$N$_\mathrm 4$ spacers are formed around the gates so that the heavily As doped source and drain are separated by approximately 70\,nm.
When negative gate voltages \VGone{} and \VGtwo{} are applied on both gates electrons are driven from source to drain through a constriction of nominal width $W$~=~30\,nm. The substrate is biased to act as a third, control back gate (with a 145\,nm buried oxide, BOX)~\cite{Roche2012}. At positive back gate voltage \VBG{} electrons in the constriction are pushed near the BOX interface. In this work we consider transport through a few implanted P donor states in this constriction (Fig.~\ref{fig1}).

\begin{figure}[t]
\begin{center}
\includegraphics[]{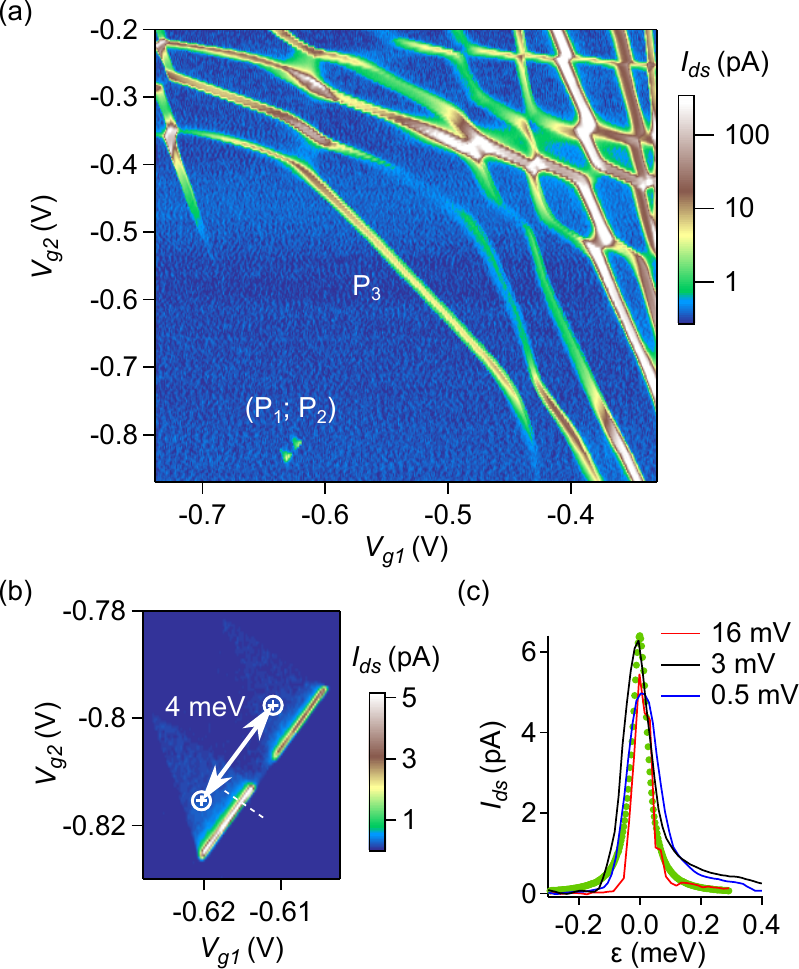}
\caption{(color online). (a) Color plot of the source-drain current versus \VGone{} and \VGtwo{} with \VD{}~=~2\,mV and \VBG~=~11.5\,V. Lines of current appear when donors---or more generally hybridized states at the bottom of the conduction band---well coupled to source and drain get ionized. Such a state is labeled P$_3$.
(\Pone{}, \Ptwo{}) corresponds to resonant current through two coupled phosphorus donors in series.
(b) Plot analogous to (a) but centered around the \Pone{}, \Ptwo{} pair. The color plot data is for \VD~=~3\,mV, the position of the triple points recorded with very small \VD{} is indicated by circles. The Coulomb repulsion between \Pone{} and \Ptwo{} is \EConetwo~=~4$\pm$0.1\,meV. (c) \IDS{} versus detuning energy across the triangle base for different \VD{}. For \VD~=~3\,mV, it corresponds to the dashed line in (b). These resonances are fitted with a Lorentzian curve (green dots, see text).}
\label{fig2}
\end{center}
\end{figure}

Figure~\ref{fig2}a shows the source-drain current \IDS{} versus \VGone{} and \VGtwo{} at fixed substrate bias \VBG{}~=~+11.5\,V, near the pinch off of the constriction. All the measurements in this paper were performed at a base temperature of 150\,mK. Lines of current appear when the energy of a donor state equals the Fermi energy in the contacts. The slope of these lines depends on the relative couplings to \VGone{} and \VGtwo{}. The antidiagonal line corresponds to the ionization (P$_3^+\rightarrow$ P$_3^0$) of a single phosphorus donor (named P$_3$) equally coupled to both front gates. The multiple anticrossings of lines visible on Figure~\ref{fig2}a are characteristic of the hybridization of two states. Approaching the onset of the conduction band, the density of states increases and the lines of current associated to the ionization of these states proliferate.

While P$_3$ gives the ionization line with the lowest energy detected,  it is not the last donor to be ionized. Below  P$_3$ an isolated double resonance emerges, which is the point where two donors {\it in series}, named \Pone{} and \Ptwo{}, get ionized. Ionization lines disappear because donors with energy well below the conduction band have less coupling with source and drain.
Resonant tunneling from source to drain through \Pone{} or \Ptwo{} is thus very weak and only tunneling via \Pone{} and \Ptwo{} in series is detectable~\cite{Larkin1987}. \Pone{} and \Ptwo{} ionization energies are 55\,meV below P$_3$. As the vertical electric field in the SOI is few mV/nm (see later on), this suggests that \Pone{} and \Ptwo{} are close to the BOX and P$_3$ is higher in the SOI.

We focus on \Pone{} and \Ptwo{}, i.e. the donors with the deepest energy levels because they are well isolated in energy from any other state. It is a necessary condition to achieve transport spectroscopy up to large energies.
The color plot of Fig.~\ref{fig2}b shows this region with a finite bias \VD{}~=~3\,mV. Two triangular shapes are clearly visible as a result of transport through the discrete levels of two dopants~\footnote{We consider only one electron per dopant. The double occupation that possibly occurs at higher energy is obscured by additional features.}.
The parameters needed to convert gate voltages into electrochemical potentials can be extracted from the shape of the triangles, which are bounded by two conditions. First the ground state of both dopants must be in the bias window, which delineates the edges. Second \Pone{}'s ground state must be higher than \Ptwo{}'s one to allow (elastic or inelastic) tunneling from the ground state of \Pone{} to \Ptwo{}, which delineates the base.
A large current flows from source to drain when a level of \Ptwo{} is resonant with the ground state of \Pone. This gives rise to lines of current at/and parallel to the base of the triangles. The sharp resonance at the triangle base corresponds to the alignment of the ground levels. The separation in energy between the \Pone{} and \Ptwo{} ground levels, called the detuning energy $\epsilon$, increases from the base ($\epsilon$~=~0) to the tip of the triangle ($\epsilon$~=~e\VD{}). The lower triangle (``electron triangle'') corresponds to an electron going from source to drain through the two ionized donors, following the sequence of charge states (0,0)$\rightarrow$(1,0)$\rightarrow$(0,1)$\rightarrow$(0,0). Transport in the upper ``hole triangle'' can be viewed as a positive charge flowing in the opposite direction through the two charged donors (1,1)$\rightarrow$(1,0)$\rightarrow$(0,1)$\rightarrow$(1,1)~\cite{Wiel2002}. The two triangles are separated by the Coulomb repulsion energy between the two localized states.

\begin{figure}
\begin{center}
\includegraphics[]{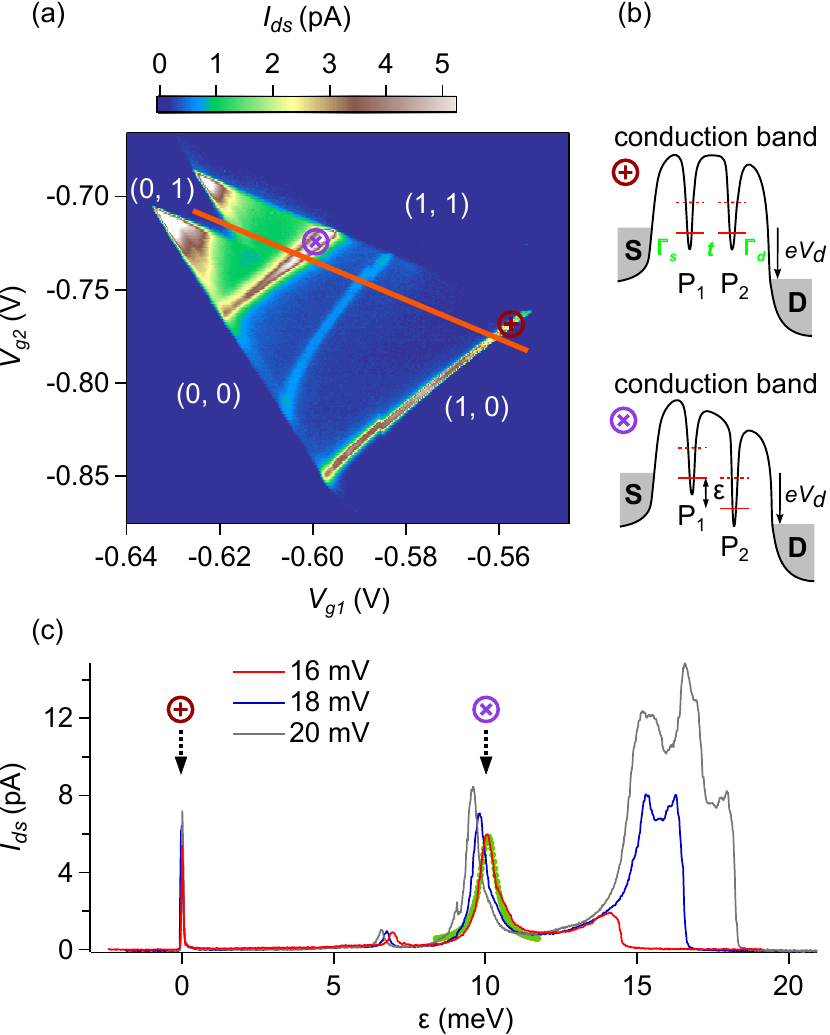}
\caption{(color online). (a) Plot analogous to Fig.~\ref{fig2}b but with \VD{}~=~16\,mV. (b) Energy diagrams (schematic) at 2 positions indicated by crosses on (a). (c) Drain current versus energy detuning $\epsilon$ between \Pone{} and \Ptwo{} states, for different \VD{}. The current is recorded along the red line in (a) for \VD{}~=~16\,mV, and the resonance with the excited state of \Ptwo{} is fitted with green dots (see text).}
\label{fig3}
\end{center}
\end{figure}

Figure~\ref{fig2}c shows \IDS{} versus $\epsilon$ for different \VD{}. A remarkable feature of resonant transport through two states in series is that the current \IDS{} should not depend on the applied voltage \VD{} (for \VD{} larger than the intrinsic energy width of the levels and temperature). For transport through two levels separated by an energy $\Delta E$, \IDS{} depends only on $\epsilon$, on the transition rates between the source and \Pone{} (\GS{}), between \Ptwo{} and the drain (\GD{}) and on the tunnel coupling between \Pone{} and \Ptwo{}, called $t$, and is given by~\cite{Stoof1996}:
\begin{equation*}
 \frac{I_{ds}}{e}= \frac{t^{2} \Gamma_d}{t^{2} (2 + \frac{\Gamma_d}{\Gamma_s})+ \frac{\Gamma_d^2}{4} + \frac{1}{\hbar^2}({\epsilon - \Delta E})^{2}}.
\label{eq1}
\end{equation*}

For the hole triangle \GD{} and \GS{} must be interchanged. Figure~\ref{fig2}c shows indeed that \IDS{} does not vary much with \VD{}. The small remaining variation can be attributed to the energy dependence of the transition rates under the electric field applied between \Pone{} and \Ptwo{}. The line shapes are well fitted by Lorentzian curves for all \VD{} and energy of the levels. Lorentzian fits near $\Delta E$~=~0 (i.e. transitions between the ground states) give \GD~$\simeq 100\,$MHz, \GS~$\simeq$ 200\,MHz $\ll$ $t$ $\simeq$ 30\,GHz. The resulting elastic current is limited by the coupling to the leads and yield to 6.5\,pA at resonance, while the width of the resonance ($\simeq$ 60\,$\mu$eV at half maximum) depends mostly on $t$.
Inelastic processes are revealed by a small, yet finite current inside the triangles, barely visible in Fig.~\ref{fig2}b. The inelastic current is rather constant for $\epsilon$ smaller than the first excited state. It can be modeled by a phenomenological tunnel rate \GIN{} which does not depend on $\epsilon$ (as in~\cite{Simmons2010}). At \VD{}~$>0$ and for $\epsilon \gg \hbar t$, \IDS{}~$\simeq e$~\GIN{} is found to be less than 0.3\,pA, setting an upper limit to \GIN{} of 2\,MHz. \GIN{}---which can be associated with phonon emission~\cite{Fujisawa1998}---is smaller than  those observed in Si/SiGe~\cite{Simmons2010}, GaAs~\cite{Fujisawa1998} and silicon quantum dots~\cite{Liu2008}.

At larger bias \VD{}~=~16\,mV, another resonance line appears within the triangles, parallel to their base for $\Delta E$~=~10.3$\pm$0.5\,meV (Fig.~\ref{fig3}). This line corresponds to the tunneling of electrons through the ground level of \Pone{} and the first excited state of \Ptwo{} as sketched in Fig.~\ref{fig3}b, and is therefore a measure of the valley-orbit splitting of \Ptwo{}. By reversing the bias voltage we probe the excited state of \Pone{} which is found to be 9.3$\pm$0.5\,meV (not shown), also very close to the bulk value. These remarkably high energies for a solid-state device is a signature of the ultimate size of the double-donor system. Previous experiments reported a VOS of only 0.1\,meV in a silicon quantum dot~\cite{Lim2011}. In two-dimensional electron gas the VOS can be higher, from fractions of a meV~\cite{Takashina2004, Goswami2007} up to $\approx 20$\,meV~\cite{Takashina2006}. The presence of uncontrolled interface states was however invoked to explain such large values~\cite{Saraiva2010}.

This spectroscopy is possible here because no other resonant state is present in the large bias energy window. Only one faint parasitic current line---not parallel to the base of the triangle---is visible in Fig.~\ref{fig3}a. This isolated line looks completely different and appears at a different energy when reversing the sign of \VD{}. It corresponds to an enhancement of the inelastic tunneling current between \Pone{} and \Ptwo{} due to a new dissipative component appearing in the nearby environment~\cite{Fujisawa1998}. It can be a resonating two level system whose one anticrossing curved branch is revealed in the bias window.

The resonance at $\Delta E$~=~10.3$\pm$0.5\,meV can also be fitted with a Lorentzian curve (Fig.~\ref{fig3}c). The maximum level of current for this resonance and for the first one are barely different. It is a strong indication that \GD{} and \GS{} are the same in both cases. Yet this resonance is broadened to 670\,$\mu$eV, eleven times larger than the resonance of the ground states. The electric field existing between the two dopants in this configuration (see Fig.~\ref{fig3}b) or the larger extent of the excited state's wave function could explain this larger linewidth though a larger coupling $t$. A short relaxation time between the excited state of \Ptwo{} and its ground state could also contribute to this broadening. Note that the inelastic current increases above the first excited state because the latter can now be involved in inelastic processes. The larger coupling $t$ between excited and ground states can explain this larger \GIN{} of 6\,MHz~\cite{Fujisawa1998}.

\begin{figure}
\begin{center}
\includegraphics[]{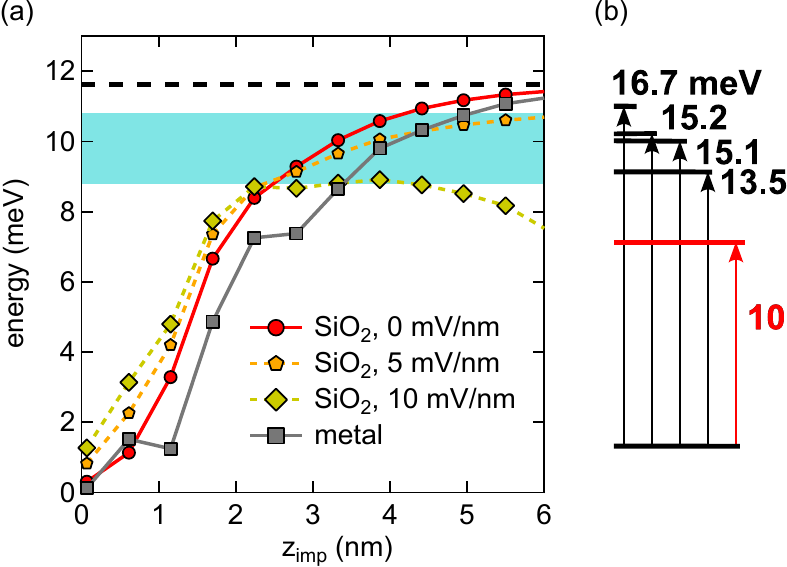}
\caption{(color online). (a) Splitting between the first two states of a P donor in a 20\,nm thick Si film embedded in SiO$_2$ as a function of the distance \ZIMP{} to the interface, for three different vertical electric fields (0, 5\,mV/nm and 10\,mV/nm). The blue interval corresponds to the measured $\Delta E$ between 8.8\,meV and 10.8\,meV for the two P donors. The splitting computed for a Si film embedded in a metal is also shown, as an upper limit to the effects of the screening by source and drain contacts.
(b) Splitting of the sixfold degenerate $1s$ state calculated for a P donor located at \ZIMP~=~3.5\,nm from the interface of a 20\,nm thick Si film embedded in SiO$_2$.}
\label{fig4}
\end{center}
\end{figure}

The measured tunnel coupling $t$ (at $\epsilon$~=~0) corresponds to a separation between the P donors ranging from 20\,nm to 30\,nm, depending on their orientation with respect to the crystal~\cite{Miller1960}.
Although $t$ can be very sensitive to the environment of the donors, this rough estimate is close both to the average distance expected between neighbor active donors ($\approx$~20\,nm) and to the distance deduced from the Coulomb repulsion between \Pone{} and \Ptwo{}. The latter, measured as \EConetwo~=~4$\pm$0.1\,meV (see Fig.~\ref{fig2}), corresponds to the bare Coulomb interaction between two electron charges 30\,nm apart in silicon.

We now focus on the small deviation of the measured VOS as compared to the bulk value. For phosphorus donors in bulk silicon the VOS is 11.67\,meV~\cite{Ramdas1981}.
To investigate how this VOS is affected by quantum confinement and electric fields~\cite{Rahman2009} we have computed the electronic structure of P impurities in a 20\,nm thick Si film embedded in SiO$_2$~\cite{Diarra2007} using a $sp^3d^5s^*$ tight-binding model~\cite{Niquet2009}.
The splitting between the two lowest impurity levels is plotted as a function of the depth \ZIMP{} of the impurity in Fig.~\ref{fig4}a. The energy levels of a P impurity located 3.5\,nm away from the lower Si/SiO$_2$ interface are shown in Fig.~\ref{fig4}b. Due to the lower symmetry of the film, the first excited state is no more degenerate and is at significantly lower energy than in bulk for \ZIMP{} near or below the Bohr radius of the impurity ($\approx$1.55\,nm). The central-cell correction is indeed reduced as the wave function is shifted away from the donor site by the dielectric interface, which decreases the valley-orbit splitting. The measured $\Delta E$ between 8.8\,meV and 10.8\,meV are actually in agreement with the calculations for P donors $\approx$\,3.5\,nm away from the Si/SiO$_2$ interface. The next excited states are then well above $\Delta E$ (Fig.~\ref{fig4}b), which is consistent with the spectroscopic data of Fig.~\ref{fig3}c. The VOS can also be reduced by the vertical electric field in the Si film and through the screening by source/drain contacts (Fig.~\ref{fig4}a). However, the effect of source and drain on the VOS is very weak for impurities at least 5\,nm from the contacts. It is unlikely that \Ptwo{} is that close to the drain given the weak \GD{}. Taking into account the source and drain bias ($V_s \approx V_d \approx 0$) and the measured values of the gate and substrate voltage at the ionization of the P donors, we estimate the electric field to be around 5\,mV/nm near the Si/BOX interface in the constriction region, which is compatible with the observation of a $\approx$\,10\,meV VOS.

In summary, we have controlled separately the energy levels of two phosphorus donors in a silicon nanowire using a compact 3-gate design. This breakthrough makes it possible to perform large bias spectroscopy and measure the VOS, which is found close to its bulk value. Deviations are in agreement with simulations for donors located 3.5\,nm above the buried oxide. By this experiment we prove that it is possible to connect electrically two isolated donors in a nano-electronic device and benefit from their very large valley splitting without significant perturbation by strong hybridization with interface states, other donors or immediate environment effects.

\begin{acknowledgments}
The authors acknowledge financial support from the EC FP7 FET-proactive NanoICT under Project AFSiD No 214989 and the French ANR under Project SIMPSSON No 2010-Blan-1015. Part of the calculations were run at the GENCI-CCRT supercomputing center.
\end{acknowledgments}

\end{document}